\documentclass[superscriptaddress,runinaddress,showpacs,showkeys,aps,prd,floatfix,onecolumn]{revtex4}
\usepackage[english]{babel}
\usepackage{amssymb,amsmath,epsfig}
\usepackage{amsmath,graphicx}
\usepackage{palatino}
\usepackage{changes}
\usepackage[colorlinks=true,
            linkcolor=blue,
            urlcolor=blue,
            citecolor=blue]{hyperref}
\begin{document}

\title{Effect of Non-linear Electrodynamics on Weak field deflection angle by Black Hole}

\author{Wajiha Javed}
\email{wajiha.javed@ue.edu.pk; wajihajaved84@yahoo.com} 
\affiliation{Division of Science and Technology, University of Education, Township-Lahore, Pakistan}

\author{Ali Hamza}
\email{alihamza.ahg@gmail.com} 
\affiliation{Division of Science and Technology, University of Education, Township-Lahore, Pakistan}

\author{Ali {\"O}vg{\"u}n}
\email{ali.ovgun@emu.edu.tr}
\affiliation{Instituto de F{\'i}sica, Pontificia Universidad Cat{\'o}lica de
Valpara{\'i}so, Casilla 4950, Valpara{\'i}so, Chile.}
\affiliation{Physics Department, Arts and Sciences Faculty, Eastern Mediterranean
University, Famagusta, North Cyprus via Mersin 10, Turkey.}
\date{\today}

\begin{abstract}
 In this work, we investigate the weak deflection angle of light from exact black hole
  within the non-linear electrodynamics. First we calculate the Gaussian optical curvature using the optical spacetime geometry. With the help of modern geometrical way popularized by 
  Gibbons and Werner, we examine the deflection angle of light from exact black hole. 
  For this desire, we determine the optical Gaussian curvature and execute 
  the Gauss-Bonnet theorem on optical metric and calculate the leading terms of deflection angle in the weak limit approximation.  Furthermore, we likewise study the 
  plasma medium's effect on weak gravitational lensing by exact black hole. Hence 
 we expose the effect of the non-linear electrodynamics on the 
  deflection angle in the weak gravitational field.
\end{abstract}



\maketitle
\section{Introduction}

In 1783, accepting the corpuscular theory of light suggested by Newton which hypothesized 
 that light comprises of small discrete particles, John Michell proposed the presence of dark stars. Michell posted a letter 
 to \textit{Philosophical affair of the Royal Society of London} \cite{C1} wherein 
 he logically said that these small discrete particles of light are
 brake by the
 star gravitational acceleration when emitted by a star, and believed that it may be 
 within reach to measure the star's mass dependent on the lighten in their rates. 
 Then again, gravitational pull of a star may be very solid even the speed of light could not escape from it, then this type of star is called dark or not visible star. Michell evaluated that might be the situation when a star having 
 size $500$ times greater than the Sun's size. Michell likewise declared that space 
 experts may identify the dark stars by studying star system gravitation-ally acting as  
 binary stars, yet here just one star could be watched. Michell's thought went ignored for 
 over a period of $100$ years, since this was accepted that gravity couldn't be associate with light.\\
Moreover, in $1915$, in theory of general relativity (GR), Einstein announced that gravitational 
 lens [dispersion of matter (for example a group of galaxies) within the light origin and 
 the viewer] could deviate the light from the light origin as the light goes near to the viewer. This impact 
 is called gravitational lensing a guess later accepted by a test \cite{C3,C4} in $1919$. A 
 gravitational lensing issue gathers a well developed theory and a wide scope of observational 
 phenomena related with the deflection of light rays with the gravity. The gravitational lens 
 theory mostly handle with geometrical optics in vacuum and uses the idea of the 
 deflection angle. The essential supposition that is the approximation of a weak deflection angle 
 of a photon. General relativity announced that a light beam going close to a circular body 
 of mass $M$ along a huge impact parameter $b$ is distracted with a little angle,
\begin{equation}
    \Theta=\frac{2R_{S}}{ b}=\frac{4M}{b},G=c=1.\label{SH}
\end{equation}
This interpretation is solid if $b\succcurlyeq R_{S}$, where $R_{S}=2M$ is the 
 Schwarzschild radius of the gravitating body. Deflection angle (\ref{SH}) is typically 
 called the "Einstein angle". In most astrophysical circumstances associated 
 with gravitational lensing, the approximation of weak deflection is well fulfilled. 
 The directions of photons in vacuum, just as the deflection angles, don't rely 
 upon the light frequency, so gravitational lensing in vacuum is neutral.\\
 It was John Wheeler who authored the word black hole (BH) and introduced the name of wormholes and contended around the idea of reality 
 with Einstein and Bohr \cite{C5}. Since $1919$, which is the time of the exploratory 
 checking of the deviation of light, various examinations on the 
 gravitational lensing have been made for the BHs as well as for the other astrophysical items  (\cite{C7}-\cite{C12}).\\
 Gravitational lensing is a helpful instrument of astrophysics \cite{D16} and astronomy, in gravitational lensing 
 light beams from distant stars and galaxies are deviate by a planet, a BH or dark matter \cite{D17,D18}. 
 The discovery of dark matter filaments \cite{D19} with the help of weak deflection is an extremely to the point topic since it is very helpful in studying the structure of the universe \cite{D20}. 
 From a hypothetical viewpoint, new techniques have been proposed to compute deflection angle. 
 In 2008, Gibbons and Werner (GW) arises with another plan to calculate the deflection angle of photon \cite{C21}. 
 Gibbons and Werner imagined that both light origin and observer lies in the asymptotic Minkowski area. 
 In the sequel,
 they utilized the Gauss-Bonnet theorem (GBT) to a spatial space, which is 
 characterized by the optical metric  \cite{C21}. In GBT, we can utilize a space $D_{R}$, 
 which is limited by the photon beam just as a circular boundary curve $C_{R}$ that is 
 situated at focus on the focal point where the photon beam meets the light origin and observer. 
 It is expected that both light origin and observer are at the coordinate length R from the 
 focal point. In the weak field approximation the GBT is stated in the form of optical metric as pursues  \cite{C21}:
\begin{equation}
    \int\int_{{D}_{R}} \mathcal{K}dS +\oint_{{\partial D}_{R}} \kappa~dt+\Sigma_{i} \theta_{i} =2\pi \mathcal{X}(D_{R}).\nonumber\\
\end{equation}
Where $\mathcal{K}$ denotes the optical Gaussian curvature and dS denotes an areal component. 
 Subsequently thinking about the Euler characteristic $\mathcal{X}(D_{R})=1$ also 
 added in the jump angles $\Sigma_{i} \theta_{i}=\pi$, the deflection angle is 
 calculated by utilizing the describing condition behaving in consistence with the 
 straight line approximation:
\begin{equation}
    \alpha=-\int_{0}^{\pi} \int_{\frac{b}{r\sin\phi}}^{\infty} \mathcal{K} dS.\nonumber\\
\end{equation}
Where deflection angle is denoted by $\Theta$. A short time later, Werner expanded this strategy 
 for stationary BHs \cite{J1 17}. Next, Ishihara et al \cite{J1 18}. demonstrated it, 
 this is achievable to calculate deflection angle for the finite distances (huge 
 impact parameter) as the GW just calculated the deflection angle 
 of BH's spacetime for the observer at asymptotically flat zone in the weak field limits 
 utilizing the optical Fermat geometry. As of late, Crisnejo and Gallo have examined the 
 deflection of light within the plasma medium \cite{J1 19}. Abdujabbarov et al. have obtained the gravitational lensing of BH in the presence of plasma \cite{pp3 1} and have also observed the effect of plasma on shadow of wormhole and BH (\cite{pp3 2}-\cite{pp3 5}). Then, Turimov et al. \cite{pp3 6} have also checked the behaviour of gravitational lensing in the presence of plasma. Moreover, Chakrabarty et al. \cite{pp3 7} and Atamurotov et al. \cite{pp3 8} have also studied the plasma medium's effect on gravitational lensing and on shadow of black hole. Moreover, Hensh et al. \cite{pp3 9} have calculated the gravitational lensing of Kehagias-Sfetsos compact objects in the presence of plasma medium.

 From that point forward, there 
 is a constantly developing concern to the weak gravitational lensing by means of the 
 strategy used by GW named as GBT for BHs, cosmic strings either wormholes (\cite{J1 20}-\cite{Li:2019mqw}). \\
 The primary point of this calculation is to explore the impact of the NLE on the 
 deflection angle of exact BH and utilize the GBT wherein the deviation of light turn 
 into a global effect. Since we just center the non-singular field outer of a light beams. 
 We mostly examined the gravitational singularities inside the general relativity. Here, 
 density clearly ends up unending at the origin of a BH and inside astronomy and 
 cosmology as the soonest condition of the cosmos during the Big Bang. In theory of (GR), 
 spacetime singularities rise various issues, both scientific and physical \cite{J1 42,J1 44}. 
 Utilizing the NLE its conceivable to resolve these singularities 
 by calculating a regular BH solution (\cite{J1 51}-\cite{J1 55}). Freshly, Kruglov suggested 
 another model of NLE with two parameters $\beta$ and $\gamma$, 
 where the particular scope of magnetic field, the unitary standards and causality are 
 fulfilled \cite{J1 41}. Furthermore, AN Aliev et al. demonstrated the impact of the 
 magnetic field on the BH spacetime \cite{J1 56,J1 57}. \\
This work is composed as pursues. In Sect. 2, we quickly survey the arrangement 
 of exact BH and after that we compute its optical metric and the 
 Gaussian optical curvature. In Sect. 3, deflection angle of light utilizing the GBT 
 is computed for exact BH. In sect. 4, we observe the graphical behaviour of deflection angle in non-plasma medium.
 In Sect. 5, we examine the effect of plasma medium on gravitational lensing. In Sect. 6, we analyzed the graphical behaviour in presence of plasma medium.
 Furthermore, we finishes up in Sect. 7 
 with a dialog in regards to the outcomes got from the present work.

\section{Exact optical metric with non-linear electrodynamics}

The action that describes the non-linear electrodynamics minimally coupled to gravity is characterized as \cite{Yu:2019xdg} follows:
\begin{equation}
 S=\frac{1}{16\pi}\int \sqrt{-g}\left(R+K(\psi)\right)d^{4}x,
\end{equation}
 where
\begin{equation*}
 \psi = {F}_{\mu \nu}  {F}^{\mu \nu} ,~~ {F}_{\mu \nu} =\nabla_\mu A_\nu -\nabla_\nu A_\mu.
\end{equation*}
Here, $R$ is Ricci scalar, $A_\mu$ is the Maxwell field, $g$ is the determinant of the metric and $K(\psi)$ 
 is defined as function of $\psi$. The field equations are calculated as follows:
\begin{equation}
 G_{\mu\nu}=-2K,_\psi F_{\mu\lambda} f^{\lambda}_{\nu} +\frac{1}{2} g_{\mu\nu} K, ~~~~ K,_\psi \equiv \frac {dK}{d\psi}
\end{equation}
 and
\begin{equation}
 \nabla_\mu(K,_\psi F^{\mu\nu})=0.
\end{equation}
In the framework of static and spherically symmetric spacetime which can generally be written as;
\begin{equation}
 ds^2=-U(r)dt^2+ \frac{dr^2}{U(r)}+r^2d\Omega^2_{2},\nonumber
\end{equation}
where $d\Omega_{2} ^{2}=d\theta^2+\sin^2\theta d\phi^2$. As the spacetime is static and spherically symmetric, so  non-vanishing $A_\mu$ is written as
\begin{equation}
 A_0=\phi(r),\nonumber
\end{equation}
and $\psi$ is
\begin{equation}
 \psi=-2\dot{\phi^{2}},\nonumber
\end{equation}
by regenerating to approximate modification of $A_\mu \rightarrow A_\mu+\nabla_\mu\mathcal{X}$. So, we get the Einstein equations and the derived Maxwell equation for the spherical symmetric spacetime
\begin{equation}
-\frac{\dot{U}\dot{f}}{f}-\frac{2U\ddot{f}}{f}+\frac{1}{f^2}-
\frac{U\dot{f^2}}{f^2}=2K_{,\psi} \dot{\phi^2}+\frac{1}{2}K,
\end{equation}
\begin{equation}
-\frac{\dot{U}\dot{f}}{f}+\frac{1}{f^2}-
\frac{U\dot{f^2}}{f^2}=2K_{,\psi} \dot{\phi^2}+\frac{1}{2}K,
\end{equation}
\begin{equation}
\frac{\dot{U}\dot{f}}{f}+
\frac{U\ddot{f}}{f}+\frac{1}{2}\ddot{U}=-\frac{1}{2}K,
\end{equation}
\begin{equation}
\left({f^2K_{,\psi}\dot{\phi}}\right)^{\textbf{.}}=0.\label{max}
\end{equation}
Dot means derivative w.r.t r. We derive these values $G^0_0=\rho$, $G^1_1=p_r$ and $G^2_2=p_\theta$
from above equations. Here Eq. \ref{max} is the equation of motion for the Maxwell field.

Presently, the measurement of static and spherical symmetric spacetime 
 with the wellspring of non-linear electrodynamics field is defined as  \cite{Yu:2019xdg}
\begin{equation}
 ds^2=-U(r)dt^2+ \frac{dr^2}{U(r)}+r^2d\Omega^2_{2},\label{AH1}
\end{equation}

 where
\begin{equation*}
 U(r)=1-\frac{2M}{r}+\frac{Q^2}{r^2}-\frac{r^2\alpha^2}{3}+2Q\alpha,
 ~~~d\Omega_{2} ^{2}=d\theta^2+\sin^2\theta d\phi^2,
\end{equation*}
 here black hole mass is denoted by $M$, charge is denoted 
 by $Q$ and $\alpha$ is the coupling constant.
 Now put the value of $U$ in Eq. (\ref{AH1}), we get the following
\begin{eqnarray}
 ds^2&=&-\left(1-\frac{2M}{r}+\frac{Q^2}{r^2}-\frac{r^2\alpha^2}{3}
 +2Q\alpha\right)dt^2+\left(1-\frac{2M}{r}+\frac{Q^2}{r^2}\right.\nonumber\\
 &-&\left.\frac{r^2\alpha^2}{3}+2Q\alpha\right)^{-1}dr^2+r^2d\theta^2+r^2\sin^2\theta d\phi^2.\label{S2}
\end{eqnarray}
By accepting that light origin and onlooker lies in the tropical plane similarly direction 
 of the null photon is additionally in a similar plane having $(\theta=\frac{\pi}{2})$.
Now, for null geodesics we put $ds^{2}$=0 and we get the following optical metric as
\begin{equation}
 dt^2=\frac{dr^2}{(1-\frac{2M}{r}+\frac{Q^2}{r^2}-\frac{r^2\alpha^2}{3}
 +2Q\alpha)^2}+\frac{r^2d\phi^2}{1-\frac{2M}{r}+\frac{Q^2}{r^2}-\frac{r^2\alpha^2}{3}+2Q\alpha}.
\end{equation}
 Now, optical metric in shape of new coordinates $r^\star$ is written as
\begin{equation}
 dt^2=\bar{g}_{ab} dx^adx^b=dr^{\star 2} +f^{2}(r^\star)d\phi^2,\label{AH2}
\end{equation}
 here
\begin{eqnarray}
 r^\star&=&\frac{r}{1-\frac{2M}{r}+\frac{Q^2}{r^2}-\frac{r^2\alpha^2}{3}+2Q\alpha},\nonumber\\
 f(r^\star)&=&\frac{r}{\sqrt{(1-\frac{2M}{r}+\frac{Q^2}{r^2}-\frac{r^2\alpha^2}{3}+2Q\alpha)}}.\label{AH3}
\end{eqnarray}
 Here we see that $(a,b)$ is converted into $(r,\phi)$ and its determinant is
 $det\bar{g}_{ab}=\frac{1}{f(r\star)^{2}}$. Now by using Eq. (\ref{AH2}),
 the non-zero christofell symbols are defined as
 \begin{equation*}
 \Gamma^{r^\star}_{\phi \phi}=-f(r^\star) f'(r^\star)~~~ and~~~
 \Gamma^\phi_{r^\star \phi}=\frac{f'(r^\star)}{f(r^\star)}
  \end{equation*}
 and the only non-vanishing Reimann tensor for optical curvature
 is given as $R_{r^\star \phi r^\star \phi}$=$-k f^{2}(r^\star)$
 where $R_{r^\star \phi r^\star \phi}$=$g_{r^\star r^\star}
 R^{r^\star}_{\phi r^\star \phi}$. Now, Gaussian optical curvature is written as
\begin{equation}
 \mathcal {K}=\frac{R_{r^\star \phi r^\star \phi}}{g_{r^\star \phi}}
= -\frac{f''(r^\star)}{f(r^\star)} =\frac{-1}{f(r^\star)} \frac{d^2 f(r^\star)}{dr^{\star2}}.
\end{equation}
 With the help of previous equation, we can say that the intrinsic Gaussian optical
 curvature denoted by $\mathcal{K}$ is written in expression of r
\begin{equation}
\mathcal {K}=\frac{-1}{f(r^\star)}\left[\frac{dr}{dr^\star}
\frac{d}{dr}(\frac{dr}{dr^\star}) \frac{df}{dr}+ \frac{d^2f}
{dr^2}(\frac{dr}{dr^\star})^2\right].\label{AH4}
\end{equation}
 Finally now we calculate the relevant Gaussian optical
 curvature for exact BH by putting the Eq. (\ref{AH3})
 into Eq. (\ref{AH4}), we calculate the suite formulation
 \begin{eqnarray}
\mathcal{K}&=&\frac{-2M}{r^3}(1-\frac{3M}{2r})+\frac{3Q^2}{r^4}
(1+\frac{2Q^2}{3r^2})-\frac{4MQ\alpha}{r^3}-\frac{6MQ^2}{r^5}\nonumber\\
&-&\frac{2Q^2\alpha}{r^2}(\alpha-\frac{3Q}{r^2})+\alpha^2(\frac{2M}{r}
-\frac{1}{3}-\frac{2Q\alpha}{3})
\end{eqnarray}
 and so, it can be written as
\begin{equation}
 \mathcal{K}=\frac{-2M}{r^3}+\frac{3Q^2}{r^4}-\frac{4MQ\alpha}{r^3}
 +\mathcal{O}(M^{-2}).\label{AH6}
\end{equation}
\section{Deflection angle of exact black hole within the non-linear electrodynamics}
  Now with the help of Gauss-Bonnet theorem we derive the deflection angle
 of a exact black hole in the presence of non-linear electrodynamics.
 We apply the Gauss-Bonnet theorem to the region $\mathcal{D}_{R}$, stated as  \cite{C21}
\begin{equation}
 \int\int_{\mathcal{D}_{R}}\mathcal{K}dS+\oint_{\partial\mathcal{D}_{R}}k dt
 +\sum_{i}\epsilon_{i}=2\pi\mathcal{X}(\mathcal{D}_{R}),
\end{equation}
 where Gaussian curvature is denoted by $\mathcal{K}$ and geodesic curvature is denoted by $k$, stated as
 $k=\bar{g}(\nabla_{\dot{\gamma}}\dot{\gamma},\ddot{\gamma})$ in such a way that $\bar{g}
 (\dot{\gamma},\dot{\gamma})=1$, here $\ddot{\gamma}$ is the representation for
 unit acceleration vector and the $\epsilon_{i}$ is the corresponding exterior angle
 at the ith vertex. As $R\rightarrow\infty$, both the jump angles become
 $\pi/2$ and we obtained $\theta_{O}+\theta_{S}\rightarrow\pi$. The Euler
 characteristic is $\mathcal{X}(\mathcal{D}_{R})=1$, as $\mathcal{D}_{R}$ is
 non singular. Therefore we get
\begin{equation}
 \int\int_{\mathcal{D}_{R}}\mathcal{K}dS+\oint_{\partial
 \mathcal{D}_{R}}kdt+\epsilon_{i}=2\pi\mathcal{X}(\mathcal{D}_{R}),
\end{equation}
 here, $\epsilon_{i}=\pi$ proves that
 $\gamma_{\bar{g}}$ and the total jump angle is a geodesic, since the Euler
 characteristic number denoted by $\mathcal{X}$ is $1$. As
 $R\rightarrow\infty$, the only interesting part to be calculated
 is $k(C_{R})=\mid\nabla_{\dot{C}_{R}}\dot{C}_{R}\mid$. Since,
 geodesic curvature$^{,}$s radial component is given by  \cite{C21}
\begin{equation}
 (\nabla_{\dot{C}_{R}}\dot{C}_{R})^{r}=\dot{C}^{\phi}_{R}
 \partial_{\phi}\dot{C}^{r}_{R}+\Gamma^{r^{\star}}_{\phi\phi}(\dot{C}^{\phi}_{R})^{2}.\label{AH5}
\end{equation}
 For large $R$, $C_{R}:=r(\phi)=R=constant$. Hence, the form
 of the equation Eq. (\ref{AH5}) becomes $(\dot{C}^{\phi}_{R})^{2}
 =\frac{1}{f^2(r^\star)}$. Remembering $\Gamma^{r^\star}_{\phi\phi}=-f(r^\star)f '(r^\star)$, it becomes
\begin{equation}
 (\nabla_{\dot{C}^{r}_{R}}\dot{C}^{r}_{R})^{r}\rightarrow\frac{1}{R}.
\end{equation}
  Hence it given that the topological defect is not involved in the geodesic curvature.
  So, $k(C_{R})\rightarrow R^{-1}$. But with the help of optical metric
 Eq. (\ref{AH2}), we can write as $dt=Rd\phi$. Hence, we came up;
\begin{equation}
 k(C_{R})dt=\frac{1}{R}Rd\phi.
\end{equation}
 Combining all the above results, we have
\begin{equation}
 \int\int_{\mathcal{D}_{R}}\mathcal{K}ds+\oint_{\partial \mathcal{D}_{R}} kdt
 =^{R \rightarrow\infty }\int\int_{S_{\infty}}\mathcal{K}dS+\int^{\pi+\Theta}_{0}d\phi.\label{hamza2}
\end{equation}
 The light ray in the weak deflection limit at 0th order is defined as $r(t)=b/\sin\phi$. So with the help of
  Eq. (\ref{AH6}) and Eq. (\ref{AH7}), the deflection angle defined as  \cite{C21}
\begin{equation}
 \Theta=-\int^{\pi}_{0}\int^{\infty}_{b/\sin\phi}\mathcal{K}\sqrt{det\bar{g}}dr^{\star}d\phi,\label{AH7}
\end{equation}
 where
\begin{equation}
 \sqrt{det\bar{g}}=r(1-\frac{3M}{r}+\frac{3Q^2}{2r^2}+3Q\alpha)dr.
\end{equation}
 After putting the leading order terms of Gaussian curvature
 Eq. (\ref{AH6}) into Eq. (\ref{AH7}), the deflection angle is defined as:
\begin{equation}
 \Theta \thickapprox \frac{4M}{b}-\frac{3\pi Q^{2}}{4b^{2}}+\frac{20MQ\alpha}{b}.\label{S1}
\end{equation}
\section{Graphical Analysis for non-plasma medium}
 This area is concerned to talk about the graphical behavior of the deflection angle.  We
 also talk about the physical importance of above mentioned plots and observe the effect of
 coupling constant $\alpha$, impact parameter $b$ and BH charge $Q$ on deflection angle.

\subsection{Deflection angle $\Theta$ w.r.t Coupling constant $\alpha$}
 
\begin{center}
\epsfig{file=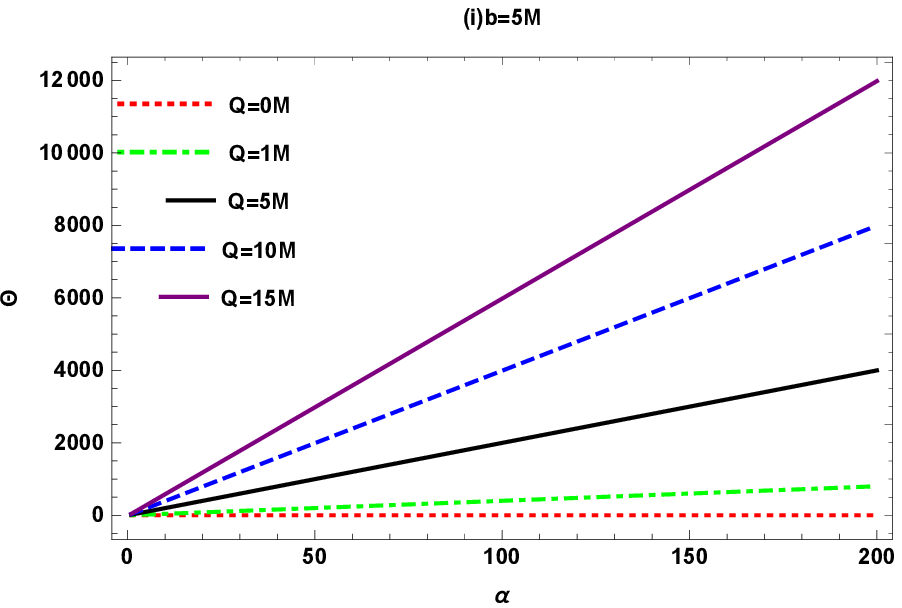,width=0.50\linewidth}\epsfig{file=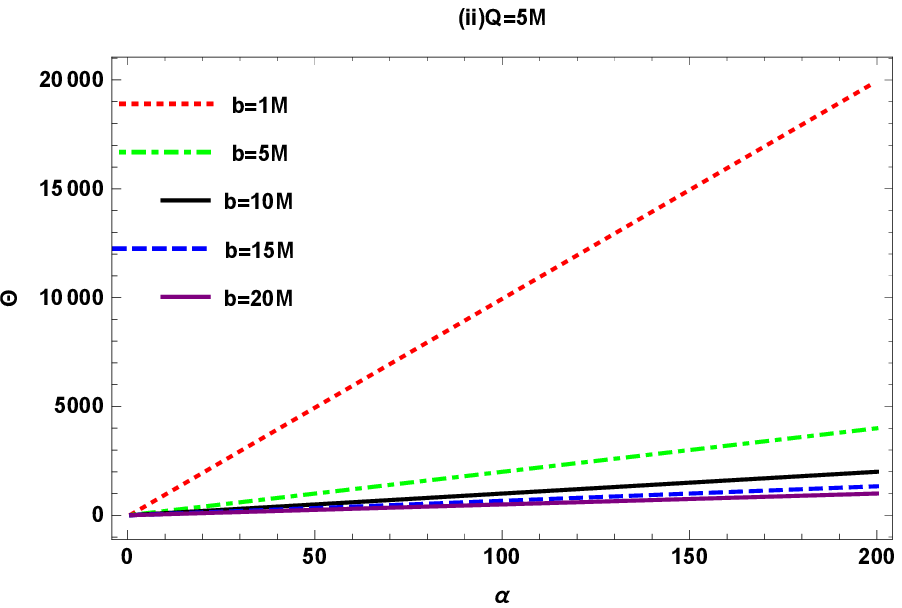,width=0.50\linewidth}\\
{Figure 1: Relation between $\Theta$ and $\alpha$}.
\end{center}
\begin{itemize}
\item \textbf{Figure 1} shows the observance of $\Theta$ w.r.t $\alpha$ by varying $Q$ and set $b=5M$ and vary the value of $b$ and fixed $Q=5M$ respectively.
\begin{enumerate}
\item In plot (i), we observed that $\Theta$ gradually decreasing for large values $Q$.
\item In plot (ii), we noted that $\Theta$ is gradually increasing for large values $b$.

\end{enumerate}
\end{itemize}
\subsection{Deflection angle $\Theta$ w.r.t Impact parameter $b$}

\begin{center}
\epsfig{file=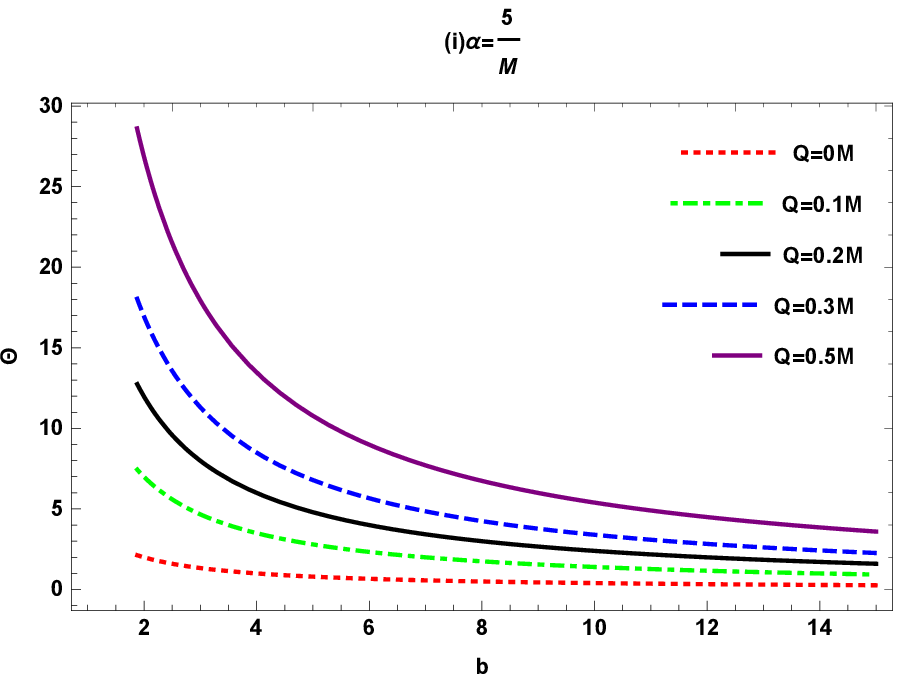,width=0.50\linewidth}\epsfig{file=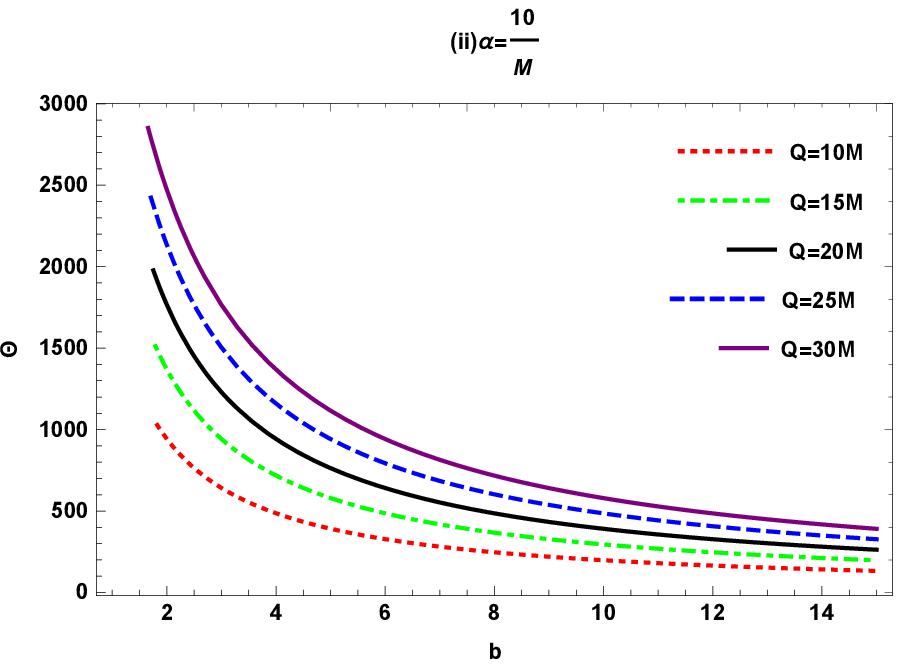,width=0.50\linewidth}\\
{Figure 2(a): Relation between $\Theta$ and $b$}.
\end{center}
\begin{itemize}
\item \textbf{Figure 2(a)} displays the observance of $\Theta$ w.r.t $b$ by taking $\alpha$ fixed and $Q$ changing.
\begin{enumerate}
\item In figure (i)and (ii), we analyzed that $\Theta$ gradually decreasing for both small and large values of $Q$.
    \end{enumerate}
\end{itemize}
  \begin{center}

\epsfig{file=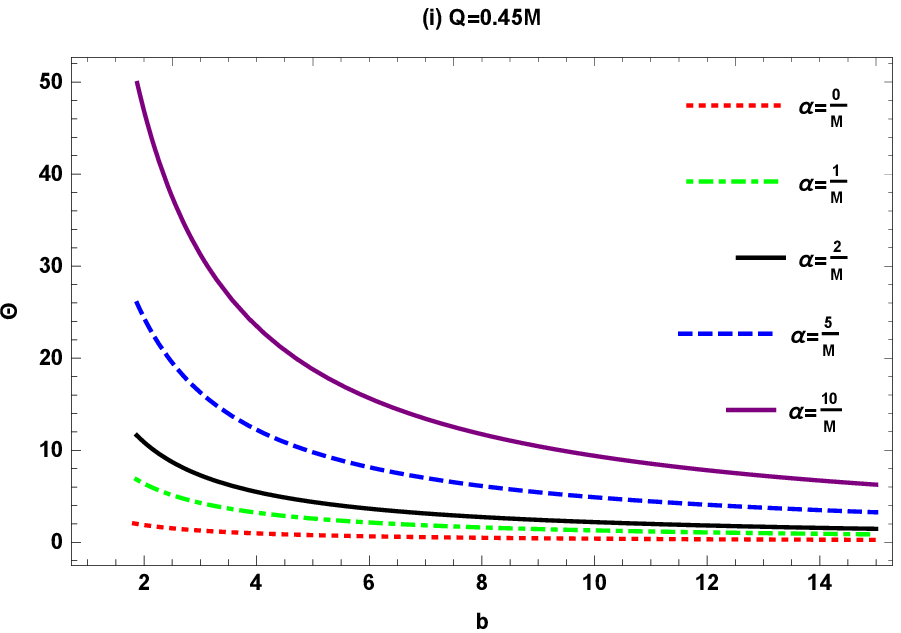,width=0.50\linewidth}\epsfig{file=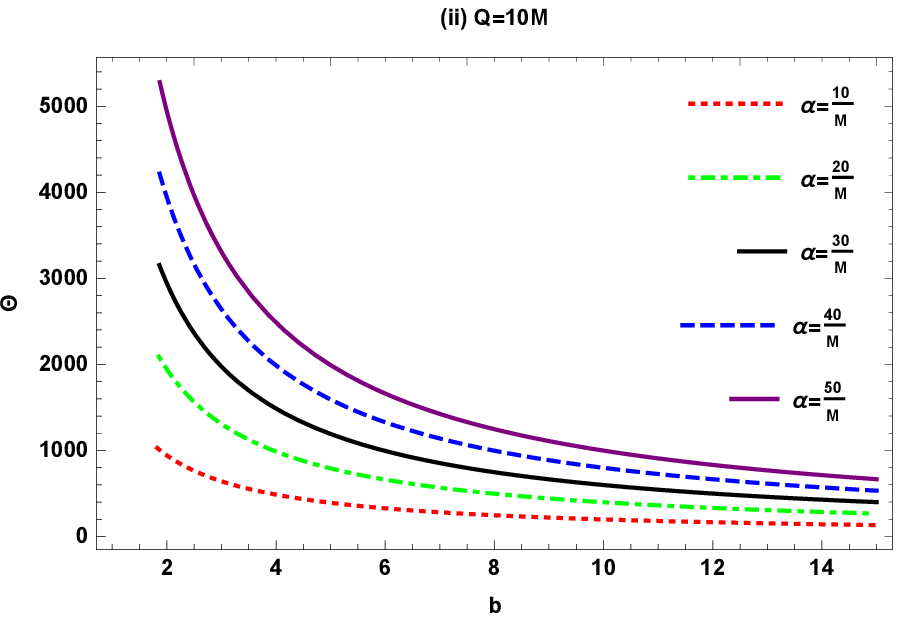,width=0.50\linewidth}\\
{Figure 2(b): Relation between $\Theta$ and $b$}.
\end{center}
\begin{itemize}
\item \textbf{Figure 2(b)} shows the observance of $\Theta$ w.r.t $b$ by taking the BH charge fixed and varying the coupling constant.
\begin{enumerate}    
\item In plot(i) and (ii), we analyze that $\Theta$ is also gradually reducing for both large and small values of coupling constant.
\end{enumerate}
\end{itemize}
\subsection{Deflection angle $\Theta$ w.r.t BH Charge $Q$}
 
\begin{center}
\epsfig{file=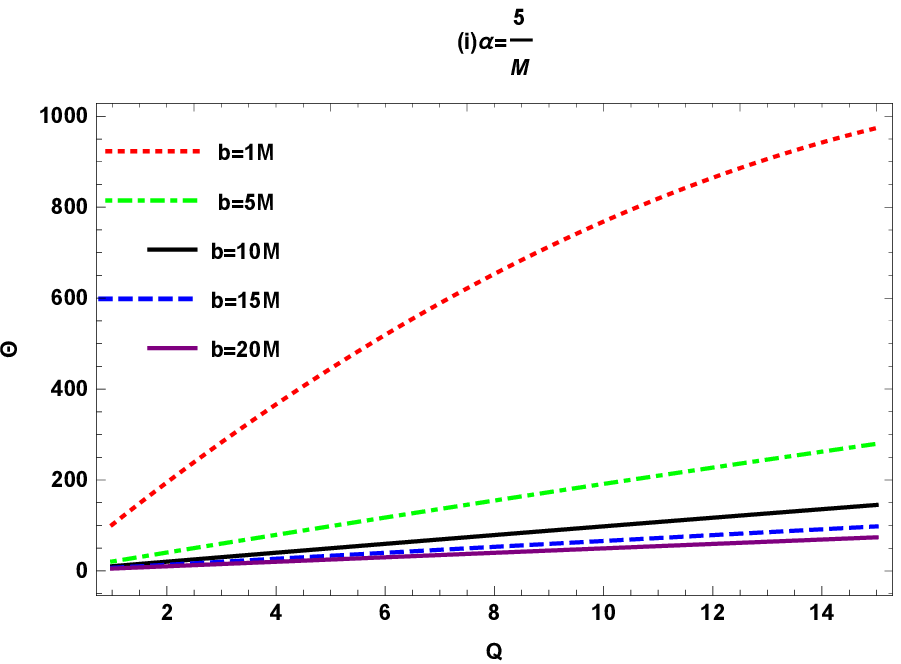,width=0.50\linewidth}\epsfig{file=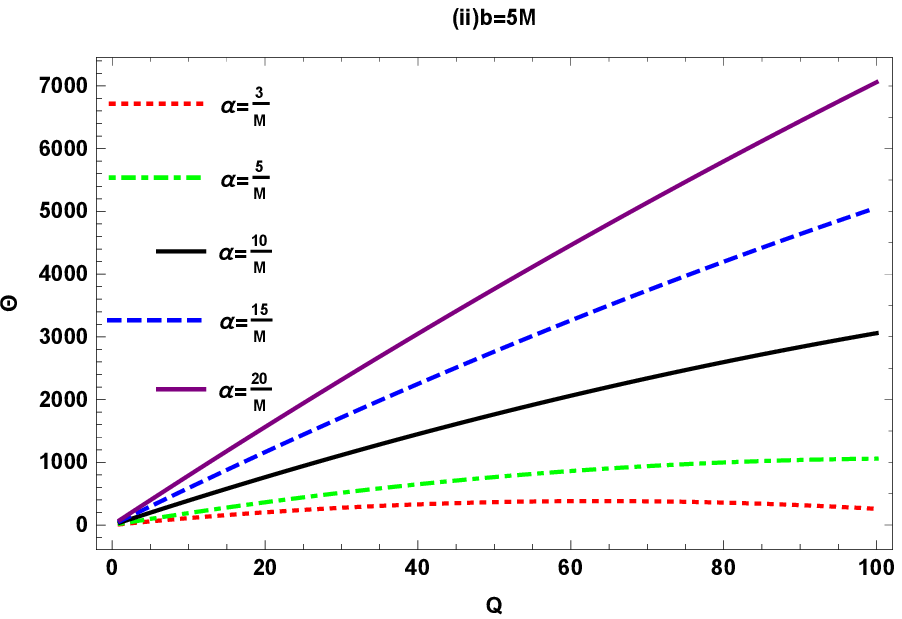,width=0.50\linewidth}\\
{Figure 3: Relation between $\Theta$ and $Q$}.
\end{center}
\begin{itemize}
\item \textbf{Figure 3} depict the observance of $\Theta$ w.r.t $Q$ for setting $\alpha=5/M$ and changing $b$ and for setting $b=5M$ and varying $\alpha$ respectively.
\begin{enumerate}
\item In picture (i), we analyzed that $\Theta$ exponentially increasing for large values of impact parameter and deflection angle rapidly increase for $1M<b<5M$.
\item In picture (ii), we examined that $\Theta$ gradually decreasing for large values of coupling constant.
\end{enumerate}
\end{itemize}

\section{Effect of plasma on gravitational lensing}
 Here, we examine the impact of plasma medium on the gravitational lensing of exact BH. Let us consider an exact BH loaded up with plasma depicted by the refractive index $n$, \cite{J1 19}
\begin{equation}
 n^2\left(r,\omega(r)\right)=1-\frac{\omega_e^2(r)}{\omega_\infty^2(r)}.
\end{equation}
 The refractive index for this case reads
\begin{equation}
 n(r)=\sqrt{{1-\frac{\omega_e^2}{\omega_\infty^2}\left(1-\frac{2M}{r}
 +\frac{Q^2}{r^2}-\frac{r^2\alpha^2}{3}+2Q\alpha\right)}},
\end{equation}
 where the metric function is defined by
\begin{equation}
 ds^2=-U(r)dt^2+ \frac{1}{U(r)}dr^2+r^2d\Omega_{2}^2
\end{equation}
 and
\begin{equation}
 U(r)=1-\frac{2M}{r}
 +\frac{Q^2}{r^2}-\frac{r^2\alpha^2}{3}+2Q\alpha.\nonumber\\
\end{equation}
 By accepting both the light origin and onlooker lies in the tropical plane similarly direction of the null photon is on a similar plane having$(\theta=\frac{\pi}{2})$.
 Now, for null geodesics we put $ds^{2}$=0 and we get the following optical metric as \cite{J1 19}
\begin{equation}
 dt^2=g^{opt}_{lm}dx^ldx^m=n^2 \left[\frac{dr^2}{U^2(r)}+\frac{r^2d\phi^2}{U(r)}\right],\label{hamza3}
\end{equation}
 with determinant $g^{opt}_{lm}$,
\begin{equation}
 \sqrt{g^{opt}}=r(1-\frac{\omega_e^2}{\omega_\infty^2})+M(3
 -\frac{\omega_e^2}{\omega_\infty^2})-\frac{Q^2}{2r}(3
 -\frac{\omega_e^2}{\omega_\infty^2})-Q\alpha r(3-\frac{\omega_e^2}{\omega_\infty^2}).
\end{equation}
 With the help of Eq. (\ref{hamza3}), we can define the non-zero christofell symbols as
\begin{equation}
    \Gamma^0_{00}=(1+\frac{\omega_e^2 A}{\omega_\infty^2})\left[-A^\prime A^{-1}(1-\frac{\omega_e^2 A}{\omega_\infty^2})-\frac{A^\prime \omega_e^2}{2\omega_\infty^2}\right],\nonumber
\end{equation}
\begin{equation}
    \Gamma_{10}^{1}=(1+\frac{\omega_e^2A}{\omega_\infty^2})\left[r^{-1}(1-\frac{\omega_e^2A}{\omega_\infty^2}-\frac{A^\prime A^{-1}}{2}(1-\frac{\omega_e^2A}{\omega_\infty^2})-\frac{A^\prime \omega_e^2}{2\omega_\infty^2}\right]\nonumber
\end{equation}
 and
\begin{equation}
    \Gamma^0_{11}=(1+\frac{A\omega_e^2}{\omega_\infty^2})\left[-rA(1-\frac{A\omega_e^2}{\omega_\infty^2})
    +\frac{r^2A^\prime}{2}(1-\frac{A\omega_e^2}{\omega_\infty^2})+\frac{r^2A}{2}\frac{A^\prime\omega_e^2}{\omega_\infty^2}\right].\nonumber
\end{equation}
 Gaussian curvature in terms of curvature tensor can be determined as
\begin{equation}
    \mathcal{K}=\frac{R_{r\phi r\phi}(g^{opt})}{det(g^{opt})},\label{hamza1}
\end{equation}
 with the help of Eq. (\ref{hamza1}) Gaussian curvature is written as
\begin{eqnarray}
\mathcal{K}&=&\frac{M}{r^3}(-2-\frac{\omega_e^2}{\omega_\infty^2}+\frac{2\omega_e^4}
{\omega_\infty^4})+\frac{2MQ^2}{r^5}(1-\frac{17\omega_e^2}{\omega_\infty^2}
+\frac{5\omega_e^4}{\omega_\infty^4})\nonumber\\ &-&\frac{4MQ\alpha}{r^3}(1+\frac{\omega_e^2}{\omega_\infty^2}
-3\frac{\omega_e^4}{\omega_\infty^4})+\mathcal{O}(M^{-2}).
\end{eqnarray}
 With the help of Gauss-Bonnet theorem we calculate the deflection angle in order to relate it with non-plasma. For calculating angle in the weak field region, as the light beams travels along a straight line approximation so  used the consideration of $ r=\frac{b}{sin\phi}$ at zero order.
\begin{equation}
    \Theta=-\lim_{R\rightarrow 0}\int_{0} ^{\pi} \int_\frac{b}{\sin\phi} ^{R} \mathcal{K} dS.
\end{equation}
 With the help of Eq. (\ref{hamza2}), the deflection angle of light in plasma medium is defined as;
\begin{equation}
    \Theta=\frac{4M}{b}-\frac{2M\omega_e^2}{b\omega_\infty^2}-\frac{6M\omega_e^4}{b\omega_\infty^4}
    -\frac{3Q^2\pi}{4b^2}+\frac{3Q^2\pi\omega_e^4}{4b^2\omega_\infty^4}+\frac{4MQ\alpha}{b}
    +\frac{2MQ\alpha\omega_e^2}{b\omega_\infty^2}.\label{S3}
\end{equation}

\section{Graphical Analysis for plasma medium}
 In this section our aim is to review the graphical behaviour of deflection angle in the presence of plasma medium. Here, we take $M$=1, $\frac{\omega_e}{\omega_\infty}$=$10^{-1}$ and vary the impact parameter, coupling constant and BH charge for obtaining these graphs.

\subsection{Deflection angle w.r.t Coupling constant}

\begin{center}
\epsfig{file=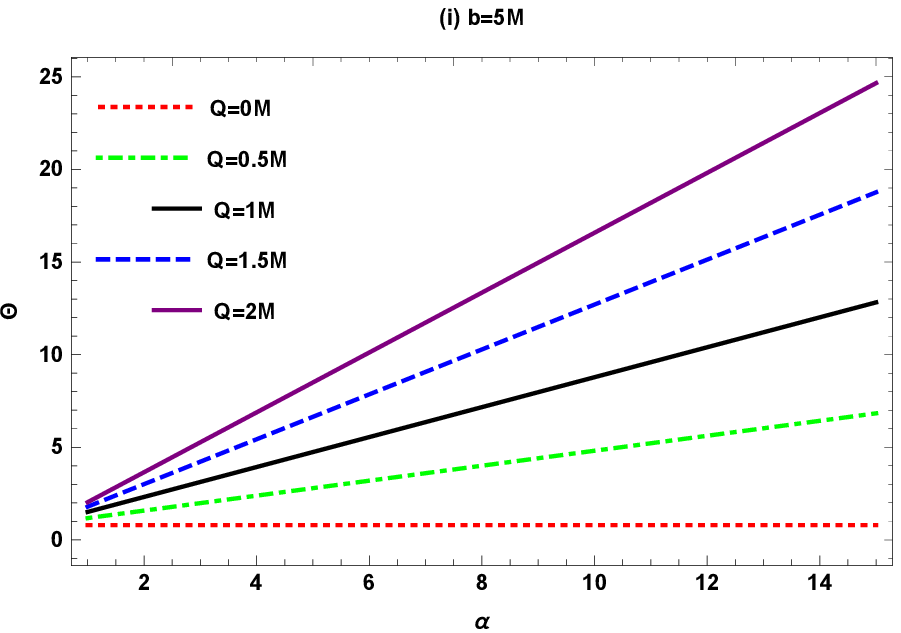,width=0.50\linewidth}\epsfig{file=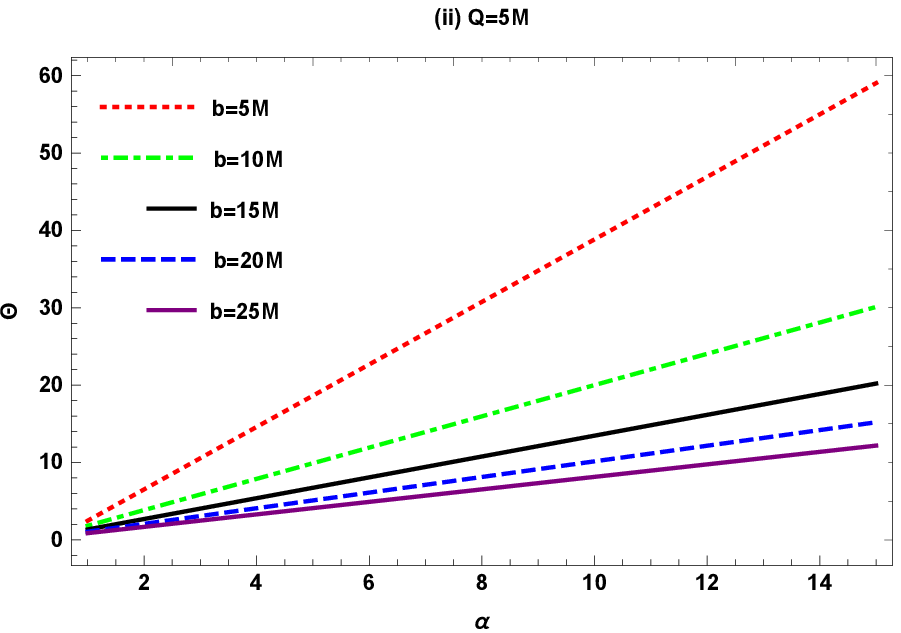,width=0.50\linewidth}\\
{Figure 4: Relation between $\Theta$ and $\alpha$}.
\end{center}

\begin{itemize}
\item \textbf{Figure 4} demonstrates the behavior of $\Theta$ w.r.t $\alpha$ for setting $b=5M$ and changing $Q$ and setting $Q=5M$ and varying $b$ respectively.
\begin{enumerate}
\item In plot (i), we saw that $\Theta$ gradually decreasing for small values of BH charge $Q$ and graph shows the positive slope.
\item In plot (ii), we analyzed that $\Theta$ exponentially increasing for large values of $b$.

\end{enumerate}
\end{itemize}

\subsection{Deflection angle $\Theta$ w.r.t Impact parameter $b$}

\begin{center}
\epsfig{file=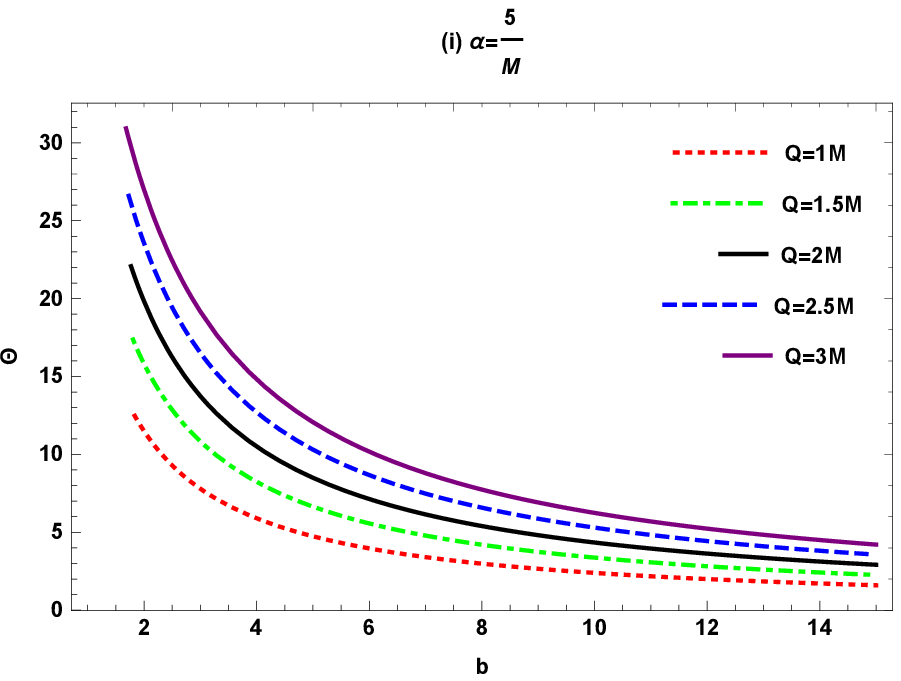,width=0.50\linewidth}\epsfig{file=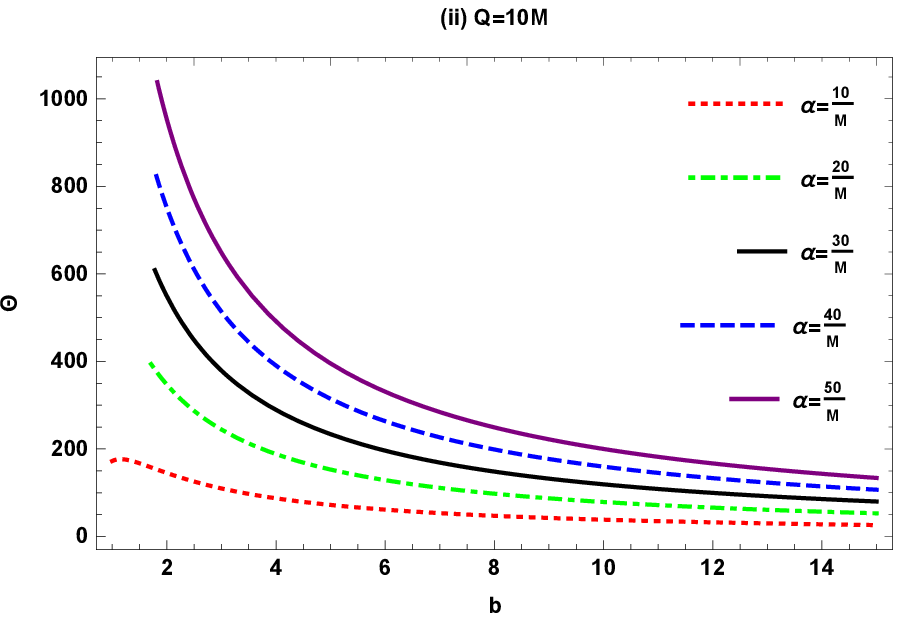,width=0.50\linewidth}\\
{Figure 5: Relation between $\Theta$ and $b$}.
\end{center}
\begin{itemize}
\item \textbf{Figure 5} shows the behaviour of $\Theta$ w.r.t $b$ for setting $\alpha=5/M$ and changing the values of $Q$ and fixed $Q$ and varying $\alpha$ respectively.
\begin{enumerate}
\item In picture (i), we analyzed that $\Theta$ gradually decreasing for low values of $Q$ and then goes to positive infinity.
\item In picture (ii), we saw that $\Theta$ is gradually decreasing for large values of $\alpha$ and then goes to positive infinity.
\end{enumerate}
\end{itemize}

\subsection{Deflection angle w.r.t Charge $Q$}
 
\begin{center}
\epsfig{file=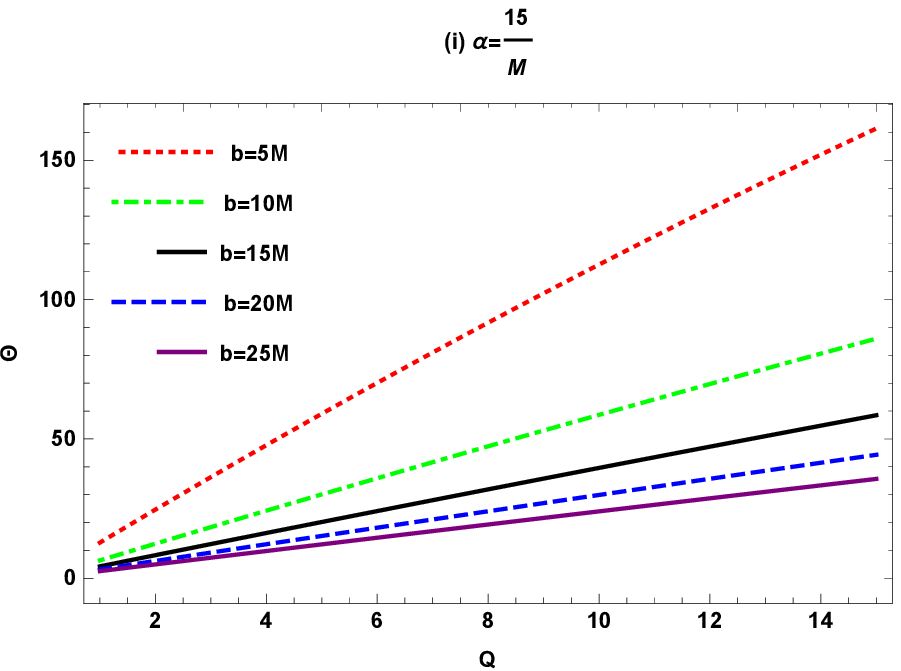,width=0.50\linewidth}\epsfig{file=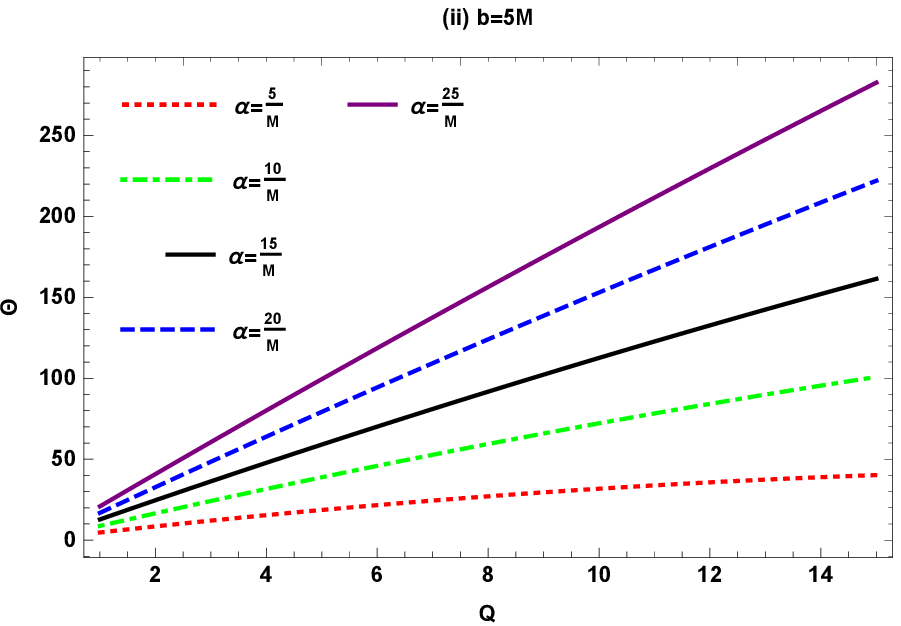,width=0.50\linewidth}\\
{Figure 6: Relation between $\Theta$ and $Q$}.
\end{center}

\begin{itemize}
\item \textbf{Figure 6} shows the behaviour of deflection angle w.r.t BH charge for fixed coupling constant and varying impact parameter and fixed impact parameter and varying coupling constant respectively.
\begin{enumerate}
\item In plot (i), we analyzed that $\Theta$ gradually increasing for large values of impact parameter and the behavior is positive slope.
\item In plot (ii), we saw that $\Theta$ gradually increasing for large values of coupling constant and the behavior is positive slope.
\end{enumerate}
\end{itemize}

\section{Summary}
 In this work, we have calculated the deflection angle for exact BH in 
 the framework of NLE. For doing this, 
  we have used the Gauss-Bonnet theorem, and we have determined the deflection angle for exact BH 
 with NLE. We have utilized the GBT and find the deflection angle of 
 photons by integrating exterior of the impact parameter, 
that represent that gravitational lensing is a global impact and is a useful asset to analyze most of the singularities of BH. 
 In this calculation, we got the deflection angle of light by exact 
 BH in the weak field limit by utilizing GBT. 
 Hence, the deflection angle (\ref{S1}) is expressed as
\begin{equation}
    \Theta \thickapprox \frac{4M}{b}-\frac{3\pi Q^{2}}{4b^{2}}
    +\frac{20MQ\alpha}{b}+\mathcal{O}(M^{-2}).\nonumber\\
\end{equation}
  By setting $Q=0$ in above equation, our proposed deflection angle 
  reduce into Schwarzschild deflection angle up to first order. 
  We hav also analyzed the graphical behavior of deflection angle for 
  exact BH in the background of NLE. Furthermore, we likewise 
  computed the deflection angle of photons by exact BH with NLE 
  in plasma medium. Deflection angle of photons in the presence of plasma medium is defined as
\begin{equation}
 \Theta=\frac{4M}{b}-\frac{2M\omega_e^2}{b\omega_\infty^2}-
 \frac{6M\omega_e^4}{b\omega_\infty^4}-\frac{3Q^2\pi}{4b^2}+\frac{3Q^2\pi\omega_e^4}{4b^2\omega_\infty^4}
    +\frac{4MQ\alpha}{b}+\frac{2MQ\alpha\omega_e^2}{b\omega_\infty^2}.\nonumber\\
\end{equation}
 By neglecting the plasma impact $(\frac{\omega_e}{\omega_\infty}\rightarrow0)$, 
 Eq. (\ref{S3}) reduce into Eq. (\ref{S1}).\\
We have observed the behaviour of deflection angle w.r.t impact
parameter b, coupling constant $\alpha$ and BH charge q.
The consequence found by analyzed the deflection angle
obtained in this paper are summed as follows:

\textit{\textbf{Deflection angle w.r.t Impact parameter:}}

 \begin{enumerate}
\item In our examination we observed that deflection angle  gradually decrease for big values of $Q$.
\item Also we examined that deflection angle is gradually decrease for big values of 
$\alpha$ which shows the stability of our proposed deflection angle.

\end{enumerate}
 \textit{\textbf{Deflection angle w.r.t Coupling constant:}}
\begin{enumerate}
\item It is to be investigated that one can only observe the stable 
behavior of deflection angle by exact BH for $0<Q\leq 2M$.
\item It is to be noted that the obtained deflection angle is positively 
increasing by increasing the impact parameter, that indicates the stable behavior.

\end{enumerate}
\textit{\textbf{Deflection angle w.r.t BH charge:}}
 \begin{enumerate}
\item We analyzed that deflection angle exponentially increasing for 
large values of impact parameter.
\item We also examined that there is direct relation between deflection angle and coupling constant.
\end{enumerate}
 To close, we have observed that in the presence of plasma medium, the deflection angle of BH decreases as compared to BH in the vacuum. As compared with these papers (\cite{pp3 1}-\cite{pp3 9}), we have also confirmed the results that deflection angle reduces more in plasma medium as compared to vacuum cases. The authors of \cite{pp3 9}, show that the radius of shadow of the black hole is increasing, when the plasma parameter increases, so that in future we will study the plasma effect on shadow of non-electrodynamics black hole. 
 \acknowledgments
This work was supported by Comisi{\'o}n Nacional de Ciencias y Tecnolog{\'i}a of Chile through FONDECYT Grant $N^\mathrm{o}$ 3170035 (A. {\"O}.).

\end{document}